\documentclass[conference]{IEEEtran}
\IEEEoverridecommandlockouts

\usepackage{cite}
\usepackage{amsmath,amssymb,amsfonts}
\usepackage{graphicx}
\usepackage{epstopdf}
\usepackage{subfigure}
\usepackage{stfloats}
\usepackage{textcomp}
\usepackage{xcolor}
\usepackage{flushend}
\usepackage{array}
\usepackage{algorithm}
\usepackage{algpseudocode}
\usepackage{multirow}
\usepackage{amsthm}

\def\BibTeX{{\rm B\kern-.05em{\sc i\kern-.025em b}\kern-.08em
    T\kern-.1667em\lower.7ex\hbox{E}\kern-.125emX}}

\begin{document}

\title{Meta-Learning-Driven GFlowNets for 3D Directional Modulation in Mobile Wireless Systems
\thanks{This work was supported by ARO grant W911NF2320103 and NSF grant ECCS-2320568.}
}

\author{\IEEEauthorblockN{Zhihao Tao}
\IEEEauthorblockA{\textit{Dept. of Electrical and Computer Engineering} \\
\textit{Rutgers, the State University of New Jersey}\\
New Brunswick, NJ, USA \\
zt118@scarletmail.rutgers.edu}
\and
\IEEEauthorblockN{Athina P. Petropulu}
\IEEEauthorblockA{\textit{Dept. of Electrical and Computer Engineering} \\
\textit{Rutgers, the State University of New Jersey}\\
New Brunswick, NJ, USA \\
athinap@soe.rutgers.edu}
}

\maketitle

\begin{abstract}
In our prior work we have proposed the use of GFlowNets, a generative AI (GenAI) framework, for designing a secure communication system comprising a
 time-modulated intelligent reflecting surface (TM-IRS).
 However,  GFlowNet-based approaches assume static environments, limiting their applicability in mobile wireless networks. 
In this paper, we proposes a novel Meta-GFlowNet framework that achieves rapid adaptation to dynamic conditions using model-agnostic meta-learning.
As the communication user is moving, the framework learns a direction-general prior across user directions via inner trajectory-balance updates and outer meta-updates, enabling quick convergence to new user directions. The approach requires no labeled data, employing a pseudo-supervised consistency objective derived from the learned reward by GFlowNet and the actual sum-rate reward of the TM-IRS system. Simulation results show that the proposed method attains faster adaptation and higher secrecy performance than retrained GFlowNets, offering an efficient GenAI framework for dynamic wireless environments.
Although the scenario considered here focuses on directional modulation–based physical-layer security, the proposed framework can also be applied to other mobile wireless systems, such as joint sensing–communication networks, that utilize GFlowNets.

\end{abstract}


\section{Introduction}


In recent years, Generative Artificial Intelligence (GenAI) has emerged as a transformative paradigm, redefining how machines can learn, create, and optimize \cite{goodfellow2014generative,kingmaauto}. From realistic image synthesis and natural language processing to complex scientific design tasks and rapidly-advancing large language models (LLMs) \cite{ho2020denoising,zhavoronkov2019deep,achiam2023gpt}, GenAI has demonstrated an unprecedented ability to model high-dimensional data distributions and generate novel, high-quality samples. Its success stems from the idea of learning generative processes, rather than purely discriminative mappings, allowing AI systems to produce diverse and coherent outputs beyond the scope of traditional supervised learning \cite{bengio2013representation,goodfellow2016deep}.

A wide spectrum of generative models has been developed, including  Generative Adversarial Networks (GANs), Variational Autoencoders (VAEs), Generative Diffusion Models (GDMs), and Generative Flow Networks (GFlowNets) \cite{khoramnejad2025generative}. 
 GFlowNets have recently gained increasing attention as a powerful alternative that unifies probabilistic sampling with sequential decision making. Unlike GANs or VAEs, which focus on matching data distributions in a single forward pass, GFlowNets model the generation of complex objects as a multi-step constructive process formulated by a Markov Decision Process (MDP) \cite{bengio2021flow}. In this framework, each trajectory represents a sequence of actions that gradually build a composite solution, and each terminal state corresponds to a complete object or configuration. The fundamental principle of GFlowNets is to train a stochastic policy such that the probability of sampling a particular terminal state is proportional to its associated reward \cite{bengio2023gflownet}. Through the flow-matching objectives, the network learns to distribute probability mass across the solution space according to a predefined reward function. This property allows GFlowNets to efficiently explore combinatorial or bounded continuous spaces, where traditional optimization techniques or reinforcement learning algorithms struggle due to sparse rewards or non-convex objectives \cite{malkin2022trajectory, tao2025time, zhang2023let}.
These advantages have recently sparked growing interest in applying GFlowNets to wireless communication and network optimization, where system parameters often form or can be converted into large discrete spaces \cite{tao2025secure, tao2025time, evmorfos2023gflownets, evmorfos2024generative, evmorfos2024gflownet, chaaya2025gflownets}. By transforming such design problems into sequential decision processes, GFlowNets enable generative and diverse sampling of high-performance parameter configurations. 

GFlowNets have been applied to sensor selection for transmit beamforming \cite{evmorfos2023gflownets} and sparse antenna array design for Integrated Sensing and Communication (ISAC) systems \cite{evmorfos2024generative, evmorfos2024gflownet}. Beyond physical-layer design, GFlowNets have also been explored for resource allocation spanning communication, sensing, and computing utilities in wireless networks \cite{chaaya2025gflownets}. Collectively, these efforts highlight GFlowNets as a promising path toward GenAI-driven optimization for next-generation intelligent wireless systems.

In our recent work \cite{tao2025secure}, we have proposed the use of GFlowNets to optimize the parameters of a time-modulated Intelligent Reflecting Surface (TM-IRS) for achieving three-dimensional (3D) directional modulation (DM)–based physical-layer security. Specifically, we considered an OFDM transmitter assisted by a TM-IRS, where each IRS element is periodically switched on and off, and its activation timing and duration define the TM-IRS parameters. These parameters are optimized to preserve signal fidelity in the desired directions toward multiple legitimate users while intentionally scrambling the signals in all other directions.
The parameter optimization problem was formulated as a deterministic MDP, in which each terminal state represents a unique TM-IRS configuration. A GFlowNet was trained to learn a stochastic policy that samples parameter configurations with probability proportional to their corresponding sum-rate rewards we formulate for the TM-IRS system, enabling efficient exploration of the high-dimensional design space. Experimental results demonstrated that the proposed approach effectively in \cite{tao2025secure} enhances the security of TM-IRS–aided OFDM systems with multiple legitimate users. Remarkably, the GFlowNet achieved convergence after training on fewer than 0.000001$\%$ of all possible configurations, highlighting its exceptional efficiency compared to exhaustive search.

However, a key limitation in our prior work and in all aforementioned works remains:  a static environment is assumed (e.g., fixed channels or user direction) during the GFlowNet training and deployment. This assumption clashes with mobile wireless operation, where user angles and propagation conditions evolve over time. One solution would be to retrain a GFlowNet whenever the environment changes; yet retraining is computationally expensive and often too slow to track variability. Another alternative is to pre-train a very large model, similar to LLMs, and subsequently fine-tune it per scenario \cite{brown2020language}. But this approach demands substantial computational and data resources beforehand, making it inefficient and impractical for problems considered here. 

In this work, we propose a meta-learning approach that equips GFlowNets with the ability to adapt rapidly to new environments with only a few updates. Specifically, we develop a Meta-GFlowNet for TM-IRS-enabled 3D directional modulation under user mobility, where each task corresponds to a desired user direction. We adopt Model-Agnostic Meta-Learning (MAML) \cite{finn2017model} to learn an initialization of GFlowNet that is broadly effective across tasks and adapts quickly to a new, unseen direction.
At a high level, our meta-training proceeds in two nested stages. In the inner loop, for a sampled direction task, we perform a small number of trajectory balance (TB) \cite{malkin2022trajectory} updates on the network parameters $\mathbf{\omega}$, which is also referred to as meta parameters, yielding task-adapted parameters $\mathbf{\varphi}$. In the outer loop, we update the meta parameters $\mathbf{\omega}$ using the TB loss computed based on the adapted parameters $\mathbf{\varphi}$, thereby learning a good prior over the task distribution. This prior captures the shared structural characteristics among user directions, allowing the GFlowNet’s flow-induced sampling distribution to align rapidly with the true objective for new directions using only a few updates. Moreover, no labeled data are required for meta-learning in this formulation. We reformulate the flow-matching objective into a pseudo-supervised consistency objective by pairing $(\Omega, R')$ with $(\Omega, R)$, where $\Omega$, $R'$, and $R$ denote the TM-IRS parameter configuration, the learned flow-implied reward, and the actual reward, respectively. The actual reward is computed analytically from the effective sum rate of the OFDM TM-IRS system in \cite{tao2025secure}. This design yields a regression-style meta-objective that teaches the model to learn to adapt its flow efficiently across diverse user directions.

The main contributions of this paper are summarized as follows:
\begin{itemize}
    \item We formulate user-direction adaptation in TM-IRS-based DM as a meta-learning problem.
    \item We develop a MAML-style Meta-GFlowNet that learns a direction-general prior and adapts to unseen directions with only a few TB updates.
    \item We demonstrate, through simulations, faster adaptation and higher secrecy performance under user motion than GFlowNets via retraining.
\end{itemize}
Although the proposed Meta-GFlowNet is developed for DM-based physical layer security, the framework can be readily extended to other GFlowNet-based optimization problems, further advancing GenAI applications in wireless systems.

\section{SYSTEM MODEL AND REWARD FORMULATION}
\begin{figure}[t]
\centerline{\includegraphics[width=3.2in]{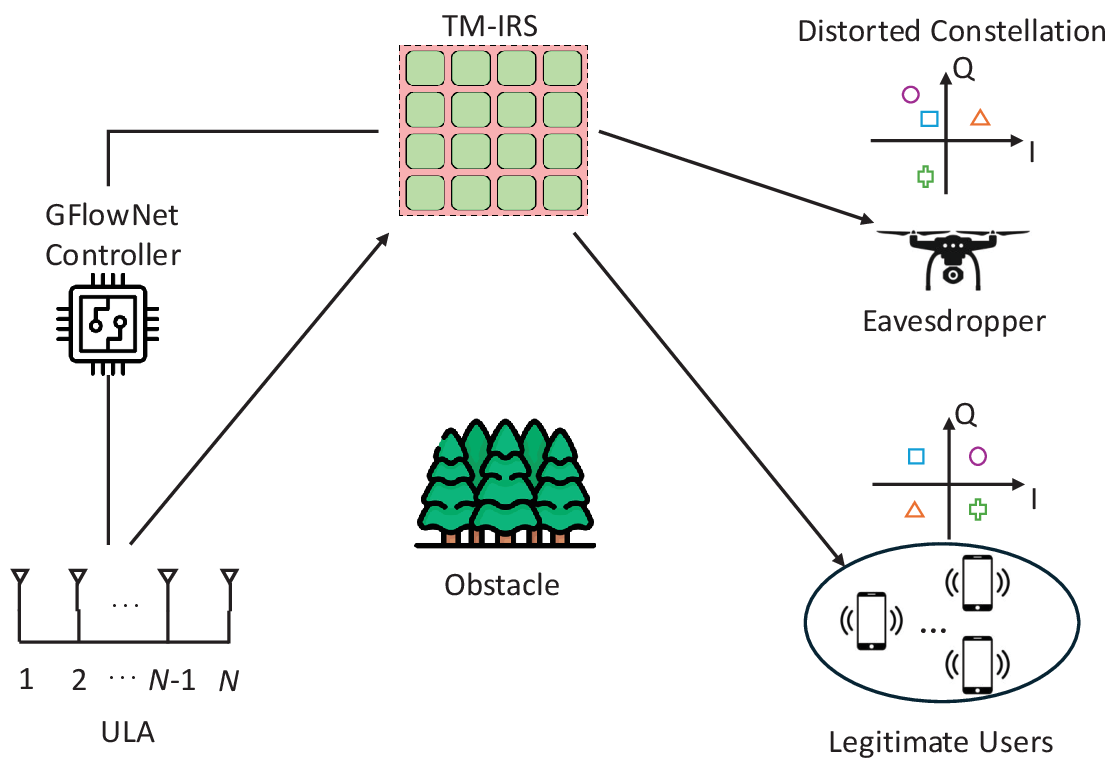}}
\caption{A TM-IRS-enabled 3D directional modulation system.}\label{fig1}
\end{figure}

We consider a TM-IRS system, aiming to achieve {3D DM security} in a {mobile wireless environment}, where the legitimate communication user (CU) is moving.
The system consists of an IRS composed of $M_x \times M_z$ passive reflecting elements that assist a {uniform linear array (ULA)} transmitter emitting {OFDM signals}, as shown in Fig.~\ref{fig1}. 
We consider the case in which the ULA communicates with the CU solely via the IRS due to obstacle blockage. 
An {eavesdropper} is also present, attempting to intercept and decode the reflected signal. 
As illustrated in Fig.~\ref{fig1}, directional modulation embeds information within the {spatial signature} of the transmitted waveform: a receiver aligned with the intended direction observes an undistorted constellation, while other directions experience scrambled symbols \cite{tao2025twc}.
Let $(\theta_T, \phi_T)$ denote the elevation and azimuth angles of the ULA transmitter relative to the IRS, and let $\theta_I$ denote the direction of the IRS as viewed from the transmitter. 
For simplicity, we consider a single legitimate user at direction $(\theta_c, \phi_c)$, while the eavesdropper is assumed static. We should note that
 a static or moving eavesdropper does not affect the proposed approach in this work, as TM-OFDM-based DM does not require knowledge of the eavesdropper’s position \cite{tao2024tma, xu2023secure}.
It is also assumed that $\theta_I$ is known at the ULA, and $(\theta_T, \phi_T)$ and $(\theta_c, \phi_c)$ are known at the IRS. 
The legitimate and eavesdropping channels are assumed known by their respective receivers, representing a {worst-case security scenario}; thus, explicit channel terms are omitted in the following for simplicity. {Neither the eavesdropper’s location nor its channel needs to be known to the ULA.}

Each ULA element transmits an OFDM signal 
$e(t)=\tfrac{1}{\sqrt{K}}\sum_{k=0}^{K-1} d(k)e^{j2\pi(f_c+kf_s)t} \; (0 \leq t < T_s)$, 
where $K$, $d(k)$, $f_c$, $f_s$, and $T_s$ denote the number of subcarriers, the modulated symbol, the carrier frequency, the subcarrier spacing, and the OFDM symbol duration, respectively. 
By applying the antenna weight $w_n=e^{-jn\pi\cos\theta_I}$ to steer the ULA beam toward the IRS, the transmitted waveform becomes 
$r(t,\theta_I)=\tfrac{1}{\sqrt{N}}\sum_{n=0}^{N-1}e(t)w_ne^{jn\pi\cos\theta_I}=\sqrt{N}\,e(t)$, 
where $N$ is the number of transmit antennas.

\subsection{Time-Modulated IRS Modeling}

Each IRS element is connected to a high-speed single-pole single-throw switch and a phase shifter. 
The switch alternates between ``on'' and ``off'' states with period $T_s$. 
Let $U_{mn}(t)$ denote the on/off switching function of the $(m,n)$-th element, with normalized turn-on time $\tau_{mn}^o \in [0,1)$ and normalized on-duration $\Delta\tau_{mn} \in [0,1)$. 
The switching function is 1 when $t \in [T_s \tau_{mn}^o,\, T_s(\tau_{mn}^o + \Delta\tau_{mn})]$ and 0 otherwise. 
Its periodic square waveform admits the Fourier expansion:
\begin{equation}\label{eq2}
\begin{split}
U_{mn}(t) = \sum_{l=-\infty}^{\infty} e^{j2\pi l f_s t} \Delta\tau_{mn}\, \mathrm{sinc}(l\pi \Delta\tau_{mn}) 
e^{-j l\pi (2\tau^o_{mn} + \Delta\tau_{mn})}.
\end{split}
\end{equation}
The IRS far-field steering vector is defined as \cite{yurduseven2020intelligent}
\begin{equation}
\begin{split}
\mathbf{a}^T(\theta, \phi) 
&= [1, e^{-j\pi \sin\theta \cos\phi}, \ldots, e^{-j\pi (M_x-1)\sin\theta \cos\phi}] \\
&\quad \otimes [1, e^{-j\pi \sin\theta \sin\phi}, \ldots, e^{-j\pi (M_z-1)\sin\theta \sin\phi}],
\end{split}
\end{equation}
where $\otimes$ denotes the Kronecker product.
Let $c_{mn}$ be the unit-modulus phase shift applied by the $(m,n)$-th element. 
The reflected signal by IRS toward direction $(\theta, \phi)$ is then
\begin{equation}\label{eq3}
y(t, \theta, \phi) = \mathbf{a}^T(\theta, \phi)\, \boldsymbol{\Phi}\, \mathbf{U}(t)\, \mathbf{a}(\theta_T, \phi_T)\, \sqrt{N}\, e(t),
\end{equation}
where $\boldsymbol{\Phi}$ and $\mathbf{U}(t)$ are diagonal matrices containing $c_{mn}$ and $U_{mn}(t)$, respectively.

Combining the above expressions we obtain
\begin{equation}
\begin{aligned}
y(t, \theta, \phi)
&= \sqrt{\frac{N}{K}} \sum_{k=0}^{K-1} d(k) e^{j2\pi (f_c + k f_s)t} \\
&\quad \times \sum_{l=-\infty}^{\infty} e^{j2\pi l f_s t} V(l, \Omega, \theta, \phi),
\end{aligned}
\end{equation}
where $\Omega = \{c_{mn}, \Delta\tau_{mn}, \tau^o_{mn}, \forall m,n\}$ denotes the TM-IRS parameter configuration, and
\begin{equation}
V(l, \Omega, \theta, \phi)
= \mathbf{a}^T(\theta, \phi)\, \boldsymbol{\Phi} \boldsymbol{\Psi}_l\, \mathbf{a}(\theta_T, \phi_T),
\end{equation}
with $\boldsymbol{\Psi}_l$ being a diagonal matrix defined as
\[
\boldsymbol{\Psi}_l =
\mathrm{diag}\!\Big(
\Delta\tau_{mn}\,
\mathrm{sinc}(l\pi \Delta\tau_{mn})\,
e^{-j l\pi (2\tau^o_{mn} + \Delta\tau_{mn})}
\Big),
\]
 capturing the harmonic-dependent amplitude and phase modulation of all IRS elements induced by time modulation.

\subsection{Reward Formulation}

After OFDM demodulation, and considering additive Gaussian noise $z_i \sim \mathcal{CN}(0, \sigma^2)$, the received symbol at subcarrier $i$ is
\begin{equation}\label{eq4}
y_i(\theta, \phi) = \sqrt{\frac{N}{K}} \sum_{k=0}^{K-1} d(k) V(i-k, \Omega, \theta, \phi) + z_i.
\end{equation}
Equation \eqref{eq4} indicates that each subcarrier is affected by weighted interference from all others, producing {inter-subcarrier scrambling}. 
Let us define $V_{i-k} = V(i-k, \Omega, \theta_c, \phi_c)$ for notational convenience. 
The signal-to-interference-plus-noise ratio (SINR) at the $i$-th subcarrier of the CU is defined as
\begin{equation}
\mathrm{SINR}_i = \frac{\eta |V_0|^2}{\eta \sum_{j=i-(K-1)}^{i} |V_j|^2 - |V_0|^2 + \sigma^2},
\end{equation}
where $\eta = N/K$. 
The achievable sum rate across all subcarriers is
\begin{equation}\label{sumC}
C_{\text{achievable}} = \sum_{i=0}^{K-1} \log_2 \big(1 + \mathrm{SINR}_i\big).
\end{equation}

To maintain constellation integrity, a phase constraint is imposed: $|\arg(V_0)| \le \xi$, where $\xi$ depends on the modulation order (e.g., $\xi < \pi/\mathcal{M}$ for $\mathcal{M}$-PSK). 
The \emph{effective sum rate} of the CU is thus defined as
\begin{equation}\label{reward}
R = C_{\text{achievable}} \cdot \mathcal{H}(\xi - |\arg(V_0)|),
\end{equation}
where $\mathcal{H}(\cdot)$ is the Heaviside step function. 
By maximizing the effective sum rate $R$ over $\Omega$,  the TM-IRS ensures an {undistorted constellation} at the legitimate user direction, while the same  is not guaranteed in undesired directions, thereby realizing {directional modulation security}. 
In the following section, we propose a {Meta-GFlowNet} design that samples high-reward $\Omega$ when the CU is moving, with the reward taken as the  $R$ of \eqref{reward}.

\section{Meta-Learning-Driven GFlowNet Framework}

In our previous work \cite{tao2025secure}, we developed a {GFlowNet-based framework} to optimize the TM-IRS parameter configuration $\Omega$ for maximizing the effective sum rate in \eqref{reward}. Specifically,
a GFlowNet models the sequential generation of a composite object as a deterministic MDP. 
Each trajectory from the root to a terminal state represents a full parameter configuration, and the goal is to learn a stochastic policy that samples terminal states with probability proportional to their reward $R$. To implement this, each TM-IRS parameter is discretized into a finite set of values, and a trajectory, denoted as $\tau=(s_0 \!\rightarrow\! s_1 \!\rightarrow\! \cdots \!\rightarrow\! s_T = \Omega)$, sequentially assigns values to all parameters until a terminal configuration $\Omega$ is reached. 
The incoming flow and outgoing flow of each state can be normalized into the forward and backward transition probabilities, $P^F$ and $P^B$, which are modeled by a neural network with learnable parameters $\mathbf{\omega}$. The model is then trained using the TB defined loss \cite{malkin2022trajectory}:
\begin{equation}\label{TBloss}
\mathcal{L}_{\mathbf{\omega}}(\tau)=
\Big(\!\log \frac{Z_\mathbf{\omega}\!\!\prod_{t=1}^{T} P_{\mathbf{\omega}}^F(s_t|s_{t-1})}
{R(s_T)\!\!\prod_{t=1}^{T} P_{\mathbf{\omega}}^B(s_{t-1}|s_t)}\!\Big)^{\!2},
\end{equation}
where \(R(s_T) = R(\Omega)\), and $Z_\mathbf{\omega}$ is the partition function, or say, the sum of rewards over all terminal states, learned from the neural network.
Minimizing \eqref{TBloss} encourages the forward policy to generate $\Omega$ with probability proportional to $R$, thus efficiently discovering high-reward solutions. 
For complete details, readers are referred to \cite{tao2025secure}.

However, this framework assumes a fixed user direction during both training and deployment, which limits its performance in {dynamic environments}. 
When the CU moves, the previously trained GFlowNet, optimized for a specific $(\theta_c,\phi_c)$, cannot generalize well to new directions. 
To overcome this limitation, we propose a meta-learning-driven GFlowNet. Meta-learning, often described as ``learning to learn'', aims to obtain a deep learning model initialization that can quickly adapt to new tasks with only a few updates. 
Here, each \emph{task} $\mathcal{T}_i$ corresponds to a specific user direction $(\theta_c^{(i)},\phi_c^{(i)})$, sampled from the region where the CU may be located, and the objective is to train a Meta-GFlowNet that generalizes across these directions.

Unlike typical supervised meta-learning that relies on large labeled datasets, our problem has no off-the-shelf data. 
However, since the GFlowNet is a generative model capable of sampling its own trajectories, we can construct \emph{synthetic self-supervised data} suitable for meta-learning. 
Specifically, for each TM-IRS configuration $\Omega$, we can analytically compute the actual reward $R$ based on \eqref{reward}. 
Meanwhile, when the flow trajectory of the GFlowNet generates the same configuration, the network yields an estimated reward, or {flow-implied reward}, $R'$, which can be derived from \eqref{TBloss} as
\begin{equation}\label{flow_reward}
    R' = \frac{Z_\mathbf{\omega}\!\!\prod_{t=1}^{T} P_{\mathbf{\omega}}^F(s_t|s_{t-1})}
{\prod_{t=1}^{T} P_{\mathbf{\omega}}^B(s_{t-1}|s_t)}.
\end{equation}
Using this mechanism, we can synthesize data pairs $(\Omega, R')$ and $(\Omega, R)$. 
For each sampled user-direction task $\mathcal{T}$, we construct two disjoint sets of sampled trajectories: a {support set} $D^{\text{sup}}_{\mathcal{T}}$ and a {query set} $D^{\text{qry}}_{\mathcal{T}}$, which contain $K_{\text{sup}}$ and $K_{\text{qry}}$ flow trajectories, respectively.
We will use these two sets in the following meta training.
This design transforms our problem into a regression-style meta-learning framework that teaches the model to align the learned flow-implied rewards $R'$ with the true analytical rewards $R$ across diverse user directions.

We adopt the MAML strategy \cite{finn2017model} to train the proposed Meta-GFlowNet via two nested stages.

\textbf{1) Inner-loop adaptation:}
For each sampled direction $\mathcal{T}_i \sim p(\mathcal{T})$, the GFlowNet parameters $\mathbf{\omega}$ are adapted over one trajectory from the support set $D^{\text{sup}}_{\mathcal{T}_i}$:
\begin{equation}\label{inner}
\mathbf{\varphi}_i = \mathbf{\omega} - \alpha \nabla_{\mathbf{\omega}} 
\mathcal{L}_{\mathbf{\omega}}^{(i)}(\tau), \; \tau \in D^{\text{sup}}_{\mathcal{T}_i}
\end{equation}
where $\alpha$ is the inner learning rate, and \(\mathcal{L}_{\mathbf{\omega}}^{(i)}(\tau)\) represents the TB loss under task $\mathcal{T}_i$.
Each trajectory in $D^{\text{sup}}_{\mathcal{T}_i}$ contributes a single gradient update to $\mathbf{\omega}$, resulting in a total of $K_{\text{sup}}$ inner-loop updates. The value of $K_{\text{sup}}$ is chosen to be significantly smaller than the number of iterations needed to train a GFlowNet from scratch, thereby substantially lowering the retraining cost.

\textbf{2) Outer-loop meta-update:}
After adaptation, the expected query loss over sampled directions is minimized, which is defined as the meta-objective, given by
\begin{equation}
\min_{\mathbf{\omega}}\;\; 
L_{\text{meta}}(\mathbf{\omega})
:= \mathbb E_{\mathcal T_i\sim p(\mathcal T)}\Big[\, 
L_{\mathbf{\varphi}_i}\big(D^{\text{qry}}_{\mathcal T_i}\big)\,\Big].
\end{equation}
The single-task query loss $L_{\mathbf{\varphi}_i}\big(D^{\text{qry}}_{\mathcal T_i}\big)$ is expressed as
\begin{equation}\label{qryloss}
 L_{\mathbf{\varphi}_i}\big(D^{\text{qry}}_{\mathcal T_i}\big) = \mathbb E_{\tau \sim D^{\text{qry}}_{\mathcal T_i}} \Big[\, \mathcal L_{\mathbf{\varphi}_i} (\tau)\,\Big].
\end{equation}
We use stochastic gradient descent to update the meta-parameters $\mathbf{\omega}$:
\begin{equation}
\mathbf{\omega} \leftarrow \mathbf{\omega} - \beta \nabla_{\mathbf{\omega}} L_{\text{meta}}(\mathbf{\omega}),
\end{equation}
where $\beta$ is the outer learning rate. The detailed training procedures are summarized in \textbf{Algorithm 1}.

\addtolength{\topmargin}{0.1in}

\begin{algorithm}[!t]
\caption{Meta-GFlowNet Training for Dynamic TM-IRS}
\begin{algorithmic}[1]
\State \textbf{Input:} Direction distribution $p(\mathcal{T})$, learning rates $\alpha$, $\beta$
\State Initialize meta-parameters $\mathbf{\omega}$
\Repeat
    \State Sample user direction $\mathcal{T}_i \sim p(\mathcal{T})$
    \For{each $\mathcal{T}_i$}
    \State Construct support set $D^{\text{sup}}_{\mathcal{T}_i}$ and query set $D^{\text{qry}}_{\mathcal{T}_i}$
    \For{each trajectory $\tau \in D^{\text{sup}}_{\mathcal{T}_i}$}
        \State Compute $P^F$, $P^B$, and $Z_\mathbf{\omega}$; derive $R'$ from \eqref{flow_reward}
        \State Compute $\mathcal{L}^{(i)}_{\mathbf{\omega}}(\tau)$ using $(\Omega, R')$ and $(\Omega, R)$
        \State Update parameters: $\mathbf{\omega} \leftarrow \mathbf{\omega} - \alpha \nabla_{\mathbf{\omega}}\mathcal{L}^{(i)}_{\mathbf{\omega}}(\tau)$
    \EndFor
    \State Let $\mathbf{\varphi}_i = \mathbf{\omega}$ and evaluate query loss using \eqref{qryloss}
    \EndFor
    \State Meta-update: $\mathbf{\omega} \leftarrow \mathbf{\omega} - \beta \nabla_{\mathbf{\omega}} L_{\text{meta}}(\mathbf{\omega})$
\Until{convergence}
\State \textbf{Output:} Meta-trained initialization $\mathbf{\omega}$
\end{algorithmic}
\end{algorithm}

Through this bi-level optimization, the Meta-GFlowNet learns a direction-general prior $\mathbf{\omega}$ that captures the shared flow structure among different user directions. During deployment, it performs only the inner-loop updates for the current direction. Via this learned prior, the deployed model generalizes efficiently to unseen directions without full retraining, allowing the flow-matching policy to quickly adapt with just a few update steps ($K_{\text{sup}}$) over the neural networks.
In contrast to a conventional GFlowNet trained for a single fixed task, the Meta-GFlowNet jointly learns both \textbf{optimization and adaptation}, enabling it to generate high-reward TM-IRS configurations on the fly. This property provide rapid responsiveness and sustained directional modulation security in mobile wireless environments. Although the Meta-GFlowNet converges more slowly than a single-task GFlowNet during training, since the former must jointly optimize over multiple tasks and approximate the task distribution, requiring more training data to converge. But the Meta-GFlowNet achieves much faster convergence during deployment as compared to retraining a GFlowNet. This is because a quick inner-loop adaptation, using only a small number of flow trajectories, are already accounted for during meta-training.

\section{Experiments}\label{simulation}
We evaluate the proposed Meta-GFlowNet in an IRS-assisted OFDM system under dynamic user motion.  
The IRS comprises $M_x \!\times\! M_z = 6\!\times\!6$ passive elements, and the transmitter employs a ULA with $N = 8$ antennas. 
The transmitted waveform contains $K = 16$ OFDM subcarriers carrying QPSK symbols. 
The transmitter is located at $(\theta_T,\phi_T) = (15^\circ,10^\circ)$, and the SNR is set to 0~dB. 
A nearest-neighbor detector is used at the receiver, and the path loss is normalized to one for simplicity. 
To accelerate training, we fix the IRS phase coefficients as 
$c_{mn} = [a_{mn}(\theta_T,\phi_T)a_{mn}(\theta_c,\phi_c)]^{-1}$ 
to steer the IRS beam always toward the legitimate user (CU), thereby excluding $c_{mn}$ from GFlowNet optimization.

\begin{figure}[t]
\centerline{\includegraphics[width=3.3in]{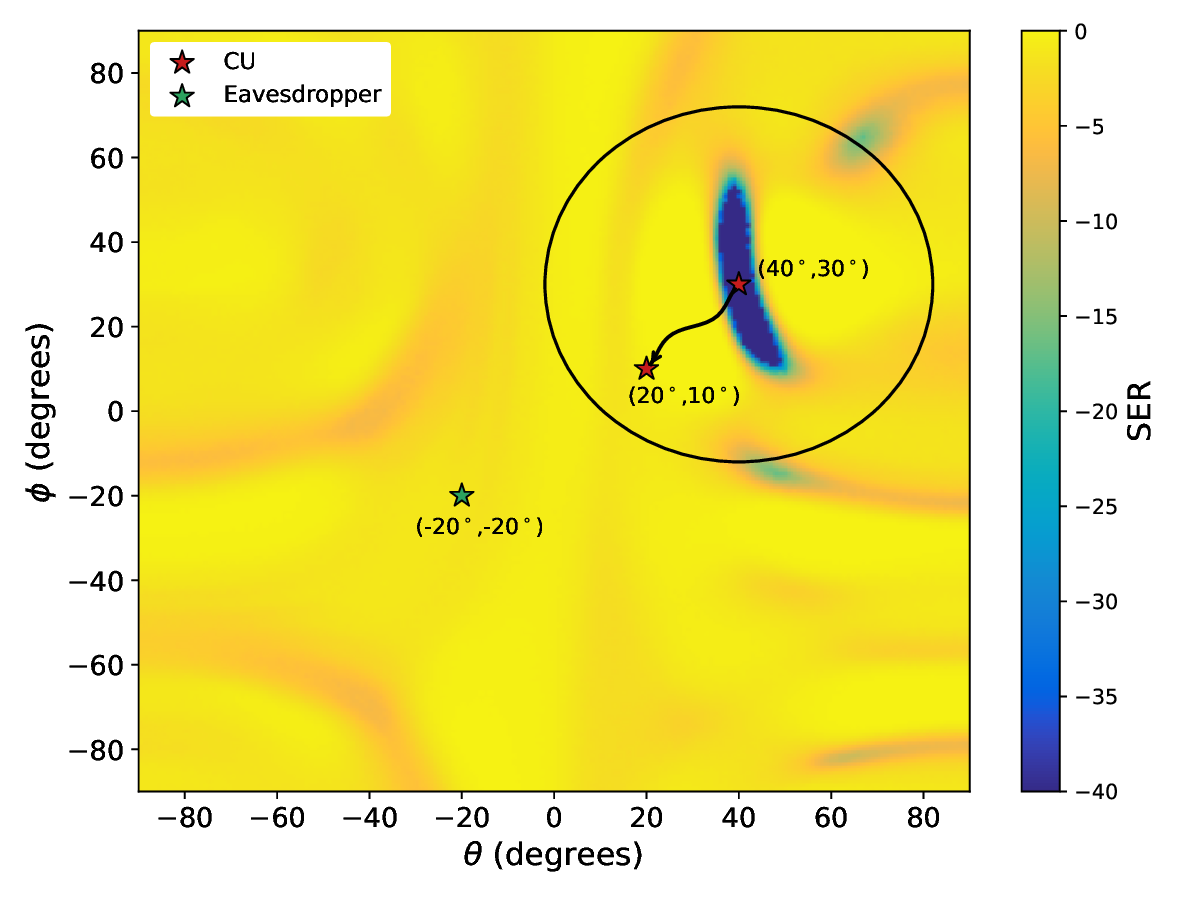}}
\caption{SER map of the conventional GFlowNet–optimized TM-IRS.}\label{fig3}
\end{figure}

Each TM-IRS parameter $\tau_{mn}^o$ and $\Delta\tau_{mn}$ is uniformly discretized into $Q = 8$ levels within $[0,1)$. 
The input state of GFlowNet $s$ is encoded as a binary vector of dimension $2M_xM_zQ\times1$, where each block corresponds to one discrete TM parameter value. 
The GFlowNet is parameterized by a multilayer perceptron (MLP) with three hidden layers of 256 neurons each, producing both forward and backward transition probabilities $(P^F,P^B)$. 
Training is conducted offline on an NVIDIA A100 GPU (32~GB memory, Google~Colab). 
For meta-training, the support and query sets of each task contain $K_{\text{sup}} = 100$ and $K_{\text{qry}} = 800$ sampled trajectories, respectively. 
The inner and meta learning rates are $\alpha = 10^{-2}$ and $\beta = 10^{-3}$, with a batch size of 10 tasks per meta-update and a total of $1\times10^6$ meta-iterations.

\subsection{Conventional GFlowNet under User Mobility}
We first examine the performance of a conventional GFlowNet (without meta-learning) when the CU moves.
As illustrated in Fig.~\ref{fig3}, the CU, denoted by a red asterisk, is initially put at $(40^\circ,30^\circ)$ and allowed to move randomly within a local region (black circle) centered at its initial position.
It is then assumed to travel toward $(20^\circ,10^\circ)$ along a curved trajectory (black line) with constant speed, representing the user’s motion path.
The eavesdropper, shown as a green asterisk, is fixed at $(-20^\circ,-20^\circ)$.
The GFlowNet is trained at the initial CU direction using $9\times10^5$ trajectories, with a learning rate of $10^{-2}$ for the first $7\times10^5$ trajectories and $10^{-3}$ for the remaining $2\times10^5$ for fine-tuning.

The learned TM-IRS parameters are then evaluated through a symbol error rate (SER) heatmap over $(\theta,\phi)\in[-90^\circ,90^\circ]$ with 1° angular resolution, as shown in Fig.~\ref{fig3}. The SER is measured in a logarithmic scale. From Fig.~\ref{fig3}, we observe that the trained model achieves a low SER around $(40^\circ,30^\circ)$, indicating the effective sum rate obtained by the learned parameters is satisfactory at that direction.
However, the SER increases rapidly as the CU moves away from the trained location, revealing that the conventional GFlowNet maintains performance only within a limited angular vicinity and lacks generalization to dynamic user mobility.

\begin{figure}[t]
\centerline{\includegraphics[width=3.1in]{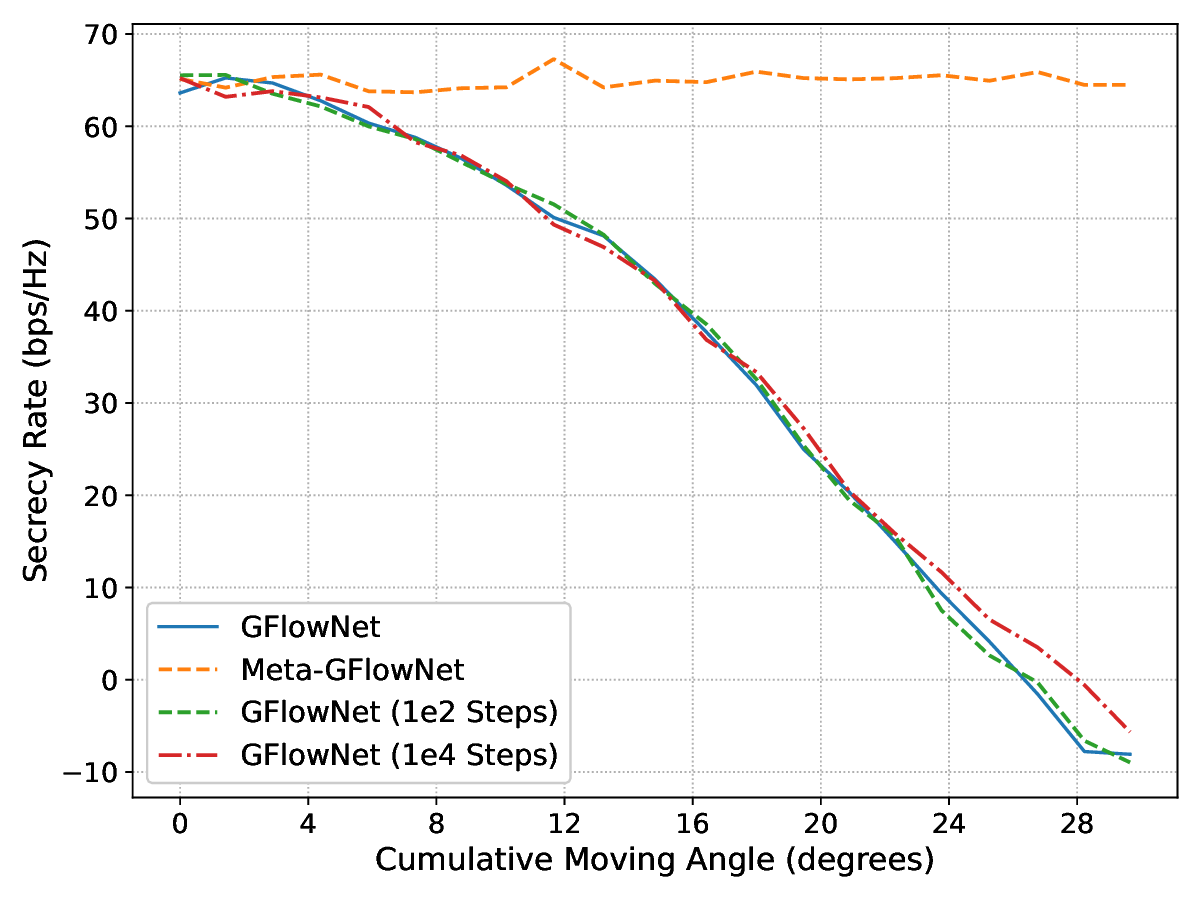}}
\caption{Secrecy-rate comparison versus cumulative moving angle among the proposed Meta-GFlowNet and benchmarks.}\label{fig5}
\end{figure}

\subsection{Performance of Meta-GFlowNet}
We now demonstrate how the proposed Meta-GFlowNet mitigates the above limitations. 
For training, each task $\mathcal{T}_i$ corresponds to a user direction uniformly sampled from the circular region shown in Fig.~\ref{fig3}. 
Meta-training follows Algorithm~1, where each task contributes $K_{\text{sup}}$ trajectories for inner-loop adaptation and $K_{\text{qry}}$ for meta-updates. 
We compare Meta-GFlowNet with three benchmarks:  
(1) the {conventional GFlowNet} trained at $(40^\circ,30^\circ)$ using $9\times10^5$ samples (denoted as \textbf{GFlowNet});  
(2) the same model retrained for $10^2$ steps (\textbf{GFlowNet (1e2~steps)}); and  
(3) retrained for $10^4$ steps (\textbf{GFlowNet (1e4~steps)}). 
All models are evaluated along the CU motion path depicted in Fig.~\ref{fig3}.

Figure~\ref{fig5} shows the secrecy rate, defined as the difference between the effective sum rates of the CU and the eavesdropper according to \eqref{reward}, plotted against the cumulative moving angle, which quantifies the total angular displacement $\sqrt{(\Delta\theta)^2 + (\Delta\phi)^2}$. 
As observed, all methods achieve comparable secrecy rates near the initial CU position (small cumulative moving angles). This similarity arises from the angular correlation of the IRS beam pattern.

However, as the CU moves farther away, the performance of the conventional GFlowNet degrades sharply. 
The model fine-tuned for $10^2$ steps shows negligible improvement over the native GFlowNet, while the $10^4$-step retrained model exhibits only marginal gains near the end of the moving path (around 24°–30° of cumulative movement). 
In contrast, the proposed Meta-GFlowNet consistently maintains high secrecy rates across the entire motion range. 
This demonstrates that the learned direction-general prior enables rapid adaptation to diverse user directions with only $K_{\text{sup}} = 100$ trajectories, whereas conventional retraining would require on the order of $9\times10^5$ samples for each new direction to achieve comparable results. 
Therefore, Meta-GFlowNet achieves stronger generalization and significantly higher adaptation efficiency, making it more suitable for dynamic wireless environments.

\section{Conclusion}
We have proposed a meta-learning–driven GFlowNet (Meta-GFlowNet) framework for optimizing time-modulated Intelligent Reflecting Surfaces (TM-IRS) in mobile wireless systems. 
The proposed Meta-GFlowNet enables fast adaptation to user mobility without retraining from scratch by learning a direction-general prior across user directions through bi-level optimization. 
Unlike typical supervised approaches, the framework requires no labeled data, leveraging pseudo-supervised reward consistency between the analytically computed and flow-implied rewards to guide training. 
Simulation results demonstrated that the Meta-GFlowNet achieves higher secrecy performance and significantly faster adaptation compared to retrained conventional GFlowNets, offering an efficient GenAI framework for secure mobile communications. 
Future work will aim to extend this framework to more dynamic wireless environments, including multi-user and joint sensing–communication scenarios, and to develop more efficient sampling mechanisms and lightweight training strategies for GFlowNets.

\bibliography{ref}

@inproceedings{xu2023secure,
title={A Secure Dual-Function Radar Communication System via Time-Modulated Arrays},
author={Xu, Zhaoyi and Petropulu, Athina},
booktitle={Proc. IEEE Radar Conference},
year={2023},
address={San Antonio, TX}
}

@inproceedings{tao2024tma,
  title={How Secure Is the Time-Modulated Array-Enabled {OFDM} Directional Modulation?},
  author={Tao, Zhihao and Xu, Zhaoyi and Petropulu, Athina},
  booktitle={Proc. IEEE Int. Conf. Acoust., Speech Signal Process. (ICASSP)},
  volume={},
  pages={},
  year={2024},
  address={Seoul, Korea}
}

@article{tao2025twc,
title={On the Security of Directional Modulation via Time Modulated Arrays Using {OFDM} Waveforms},
author={Tao, Zhihao and Petropulu, Athina},
journal={IEEE Tran. on Wire. Commun.},
note={to appear},
year={2025},
url={https://arxiv.org/abs/2408.10522}}

@article{bengio2021flow,
  title={Flow network based generative models for non-iterative diverse candidate generation},
  author={Bengio, Emmanuel and Jain, Moksh and Korablyov, Maksym and Precup, Doina and Bengio, Yoshua},
  journal={Advances in Neural Information Processing Systems},
  volume={34},
  pages={27381--27394},
  year={2021}
}

@article{bengio2023gflownet,
  title={Gflownet foundations},
  author={Bengio, Yoshua and Lahlou, Salem and Deleu, Tristan and Hu, Edward J and Tiwari, Mo and Bengio, Emmanuel},
  journal={Journal of Machine Learning Research},
  volume={24},
  number={210},
  pages={1--55},
  year={2023}
}

@article{yurduseven2020intelligent,
  title={Intelligent reflecting surfaces with spatial modulation: An electromagnetic perspective},
  author={Yurduseven, Okan and Assimonis, Stylianos D and Matthaiou, Michail},
  journal={IEEE Open Journal of the Communications Society},
  volume={1},
  pages={1256--1266},
  year={2020},
  publisher={IEEE}
}

@article{malkin2022trajectory,
  title={Trajectory balance: {Improved} credit assignment in gflownets},
  author={Malkin, Nikolay and Jain, Moksh and Bengio, Emmanuel and Sun, Chen and Bengio, Yoshua},
  journal={Advances in Neural Information Processing Systems},
  volume={35},
  pages={5955--5967},
  year={2022}
}

@INPROCEEDINGS{tao2025secure,
  title={Secure Time-Modulated Intelligent Reflecting Surface via Generative Flow Networks},
  author={Tao, Zhihao and Petropulu, Athina P},
  booktitle={Proc. of the IEEE MILCOM 2025},
  address={Los Angeles, CA, USA},
  year={2025}
}

@article{tao2025time,
  title={Time-Modulated Intelligent Reflecting Surfaces for Integrated Sensing, Communication and Security: {A} Generative {AI} Design Framework},
  author={Tao, Zhihao and Petropulu, Athina and Poor, H Vincent},
  journal={arXiv preprint arXiv:2509.05565},
  year={2025}
}

@inproceedings{evmorfos2023gflownets,
  title={Gflownets for sensor selection},
  author={Evmorfos, Spilios and Xu, Zhaoyi and Petropulu, Athina},
  booktitle={2023 IEEE 33rd International Workshop on Machine Learning for Signal Processing (MLSP)},
  pages={},
  year={2023}
}

@inproceedings{evmorfos2024generative,
  title={Generative {AI} for Sparse Antenna Array Design in {ISAC} Systems},
  author={Evmorfos, Spilios and Petropulu, Athina P},
  booktitle={2024 IEEE 25th International Workshop on Signal Processing Advances in Wireless Communications (SPAWC)},
  pages={306--310},
  year={2024}
}

@inproceedings{evmorfos2024gflownet,
  title={GFlowNet-Based Antenna Selection for {ISAC} Systems Under the Presence of Eavesdroppers},
  author={Evmorfos, Spilios and Petropulu, Athina P},
  booktitle={2024 58th Asilomar Conference on Signals, Systems, and Computers},
  pages={438--442},
  year={2024}
}

@article{chaaya2025gflownets,
  title={GFlowNets for Active Learning Based Resource Allocation in Next Generation Wireless Networks},
  author={Chaaya, Charbel Bou and Bennis, Mehdi},
  journal={arXiv preprint arXiv:2505.05224},
  year={2025}
}

@inproceedings{finn2017model,
  title={Model-agnostic meta-learning for fast adaptation of deep networks},
  author={Finn, Chelsea and Abbeel, Pieter and Levine, Sergey},
  booktitle={International conference on machine learning (ICML)},
  pages={1126--1135},
  year={2017}
}

@article{brown2020language,
  title={Language models are few-shot learners},
  author={Brown, Tom and Mann, Benjamin and Ryder, Nick and Subbiah, Melanie and Kaplan, Jared D and Dhariwal, Prafulla and Neelakantan, Arvind and Shyam, Pranav and Sastry, Girish and Askell, Amanda and others},
  journal={Advances in neural information processing systems (NeurIPS)},
  volume={33},
  pages={1877--1901},
  year={2020}
}

@article{khoramnejad2025generative,
  title={Generative {AI} for the optimization of next-generation wireless networks: Basics, state-of-the-art, and open challenges},
  author={Khoramnejad, Fahime and Hossain, Ekram},
  journal={IEEE Communications Surveys \& Tutorials},
  year={2025},
  publisher={IEEE}
}

@article{goodfellow2014generative,
  title={Generative adversarial nets},
  author={Goodfellow, Ian J and Pouget-Abadie, Jean and Mirza, Mehdi and Xu, Bing and Warde-Farley, David and Ozair, Sherjil and Courville, Aaron and Bengio, Yoshua},
  journal={Advances in neural information processing systems (NeurIPS)},
  volume={27},
  year={2014}
}

@inproceedings{kingmaauto,
  title={Auto-encoding variational {Bayes}},
  author={Kingma, Diederik P and Welling, Max},
  booktitle={Int. Conf. on Learning Representations (ICLR)},
  year={2014}
}

@article{ho2020denoising,
  title={Denoising diffusion probabilistic models},
  author={Ho, Jonathan and Jain, Ajay and Abbeel, Pieter},
  journal={Advances in neural information processing systems (NeurIPS)},
  volume={33},
  pages={6840--6851},
  year={2020}
}

@article{zhavoronkov2019deep,
  title={Deep learning enables rapid identification of potent {DDR1} kinase inhibitors},
  author={Zhavoronkov, Alex and Ivanenkov, Yan A and Aliper, Alex and Veselov, Mark S and Aladinskiy, Vladimir A and Aladinskaya, Anastasiya V and Terentiev, Victor A and Polykovskiy, Daniil A and Kuznetsov, Maksim D and Asadulaev, Arip and others},
  journal={Nature biotechnology},
  volume={37},
  number={9},
  pages={1038--1040},
  year={2019},
  publisher={Nature Publishing Group US New York}
}

@article{achiam2023gpt,
  title={{GPT}-4 technical report},
  author={Achiam, Josh and Adler, Steven and Agarwal, Sandhini and Ahmad, Lama and Akkaya, Ilge and Aleman, Florencia Leoni and Almeida, Diogo and Altenschmidt, Janko and Altman, Sam and Anadkat, Shyamal and others},
  journal={arXiv preprint arXiv:2303.08774},
  year={2023}
}

@article{bengio2013representation,
  title={Representation learning: A review and new perspectives},
  author={Bengio, Yoshua and Courville, Aaron and Vincent, Pascal},
  journal={IEEE transactions on pattern analysis and machine intelligence},
  volume={35},
  number={8},
  pages={1798--1828},
  year={2013},
  publisher={IEEE}
}

@book{goodfellow2016deep,
  title={Deep learning},
  author={Goodfellow, Ian and Bengio, Yoshua and Courville, Aaron and Bengio, Yoshua},
  volume={1},
  number={2},
  year={2016},
  publisher={MIT press Cambridge}
}

@article{zhang2023let,
  title={Let the flows tell: {Solving} graph combinatorial problems with gflownets},
  author={Zhang, Dinghuai and Dai, Hanjun and Malkin, Nikolay and Courville, Aaron C and Bengio, Yoshua and Pan, Ling},
  journal={Advances in neural information processing systems (NeurIPS)},
  volume={36},
  pages={11952--11969},
  year={2023}
}
\bibliographystyle{IEEEtran}

\end{document}